\documentstyle[11pt,newpasp,twoside,epsf]{article}

\begin{document}

\title{The Shape of the Milky Way Halo and the Satellite Tidal Tails}
\author{M.A. G\'omez-Flechoso}
\affil{Universidad Europea de Madrid, Dpto. de Matem\'aticas, E-28670
Villaviciosa de Od\'on (Spain)}

\begin{abstract}

The dwarf galaxies orbiting a main galaxy suffer strong tidal forces
produced by its dark halo. As a consequence,
substructures and tidal tails could appear in the satellites.
These structures
could give us information about the dark matter content of the main
and the dwarf galaxies.
The Milky Way satellites, because of their proximity, are a good sample
to study the effects of the tidal forces.
The shape of the Milky Way potential could be inferred from the
observational data of the tidal tails of its satellites.
The Sagittarius dwarf is one of the most interesting satellites as
it presents a long tidal tail that covers a wide angle on the sky with a
large variation of heliocentric distance along the stream.
\end{abstract}

\section{Introduction}

   Several processes are involved in the interaction between
a galaxy and its dwarf satellites. The most important are
the dynamical friction (producing the satellite orbital decay),
the disk shocking (that heats the disc and disrupts the satellite),
the variation of the primary galaxy potential by the mass accretion
of the satellites, and the tidal stripping.

In this paper, we will study the tidal stripping.
It gives place to the formation of two almost equal tidal tails, the
leading and trailing tails, that roughly follow the satellite orbit.
As the potential is the physical quantity that determines this
orbit, if the shape of the tidal tails is observed, the main galaxy potential can be
inferred.

The tidal streams of the satellite galaxies have been observed in some external
galaxies as polar rings 
or traces of tidal tails. 
They have also been detected in some globular clusters
and dwarf galaxies
of the Milky Way (MW).

\section{The tidal streams and the Milky Way system}

Among the MW satellite galaxies having tidal tails, the most prominent
is the Sagittarius (Sgr) dwarf spheroidal galaxy (Ibata et al. 1994).
A great circle of almost 360 degrees has been recently
identified as Sgr tidal stream (Majewski et al. 2003). Besides that, a great
number of observations of the Sgr stream has been reported (Mateo et al. 1998;
Majewski et al. 1999; Ivezi\'c et al. 2000;  Yanny et al. 2000; Ibata et al. 2001a; Mart\'{\i}nez-Delgado
et al. 2001; Dinescu et al. 2002; Mart\'{\i}nez-Delgado et al. 2004). This set of
observations provides data about the projected position, the heliocentric distance
and the radial velocity of the stars in several parts of the stream. It
gives us information about the shape and the kinematics of the Sgr orbit (see
Sect. 2.2 and Fig. 1) and, therefore, about the shape of the MW potential.
The importance of the Sgr stream is that it maps the whole orbit, covering a
wide range of galactocentric distances and travelling through inner and
outer regions of the MW potential.

\subsection{The Sagittarius + Milky Way model}

We have used a three component model (G\'omez-Flechoso
et al. 1999) with several flatness parameters for the halo density, $q_d$,
to describe the MW.
In this potential, we have calculated the orbit of Sgr, forcing the model
to reproduce the present position and velocity of Sgr.

\subsection{Results}

\begin{figure}
\plottwo{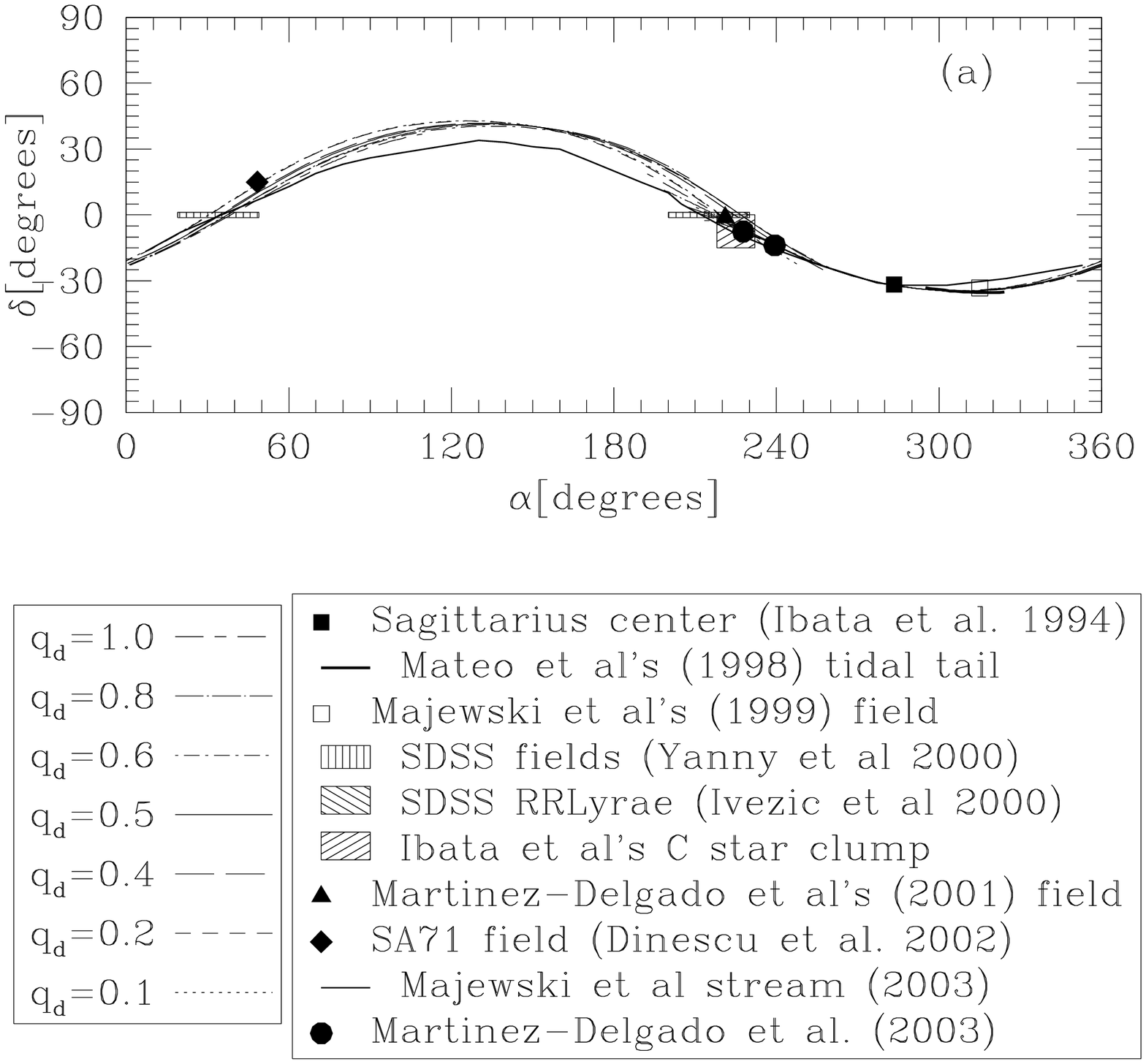}{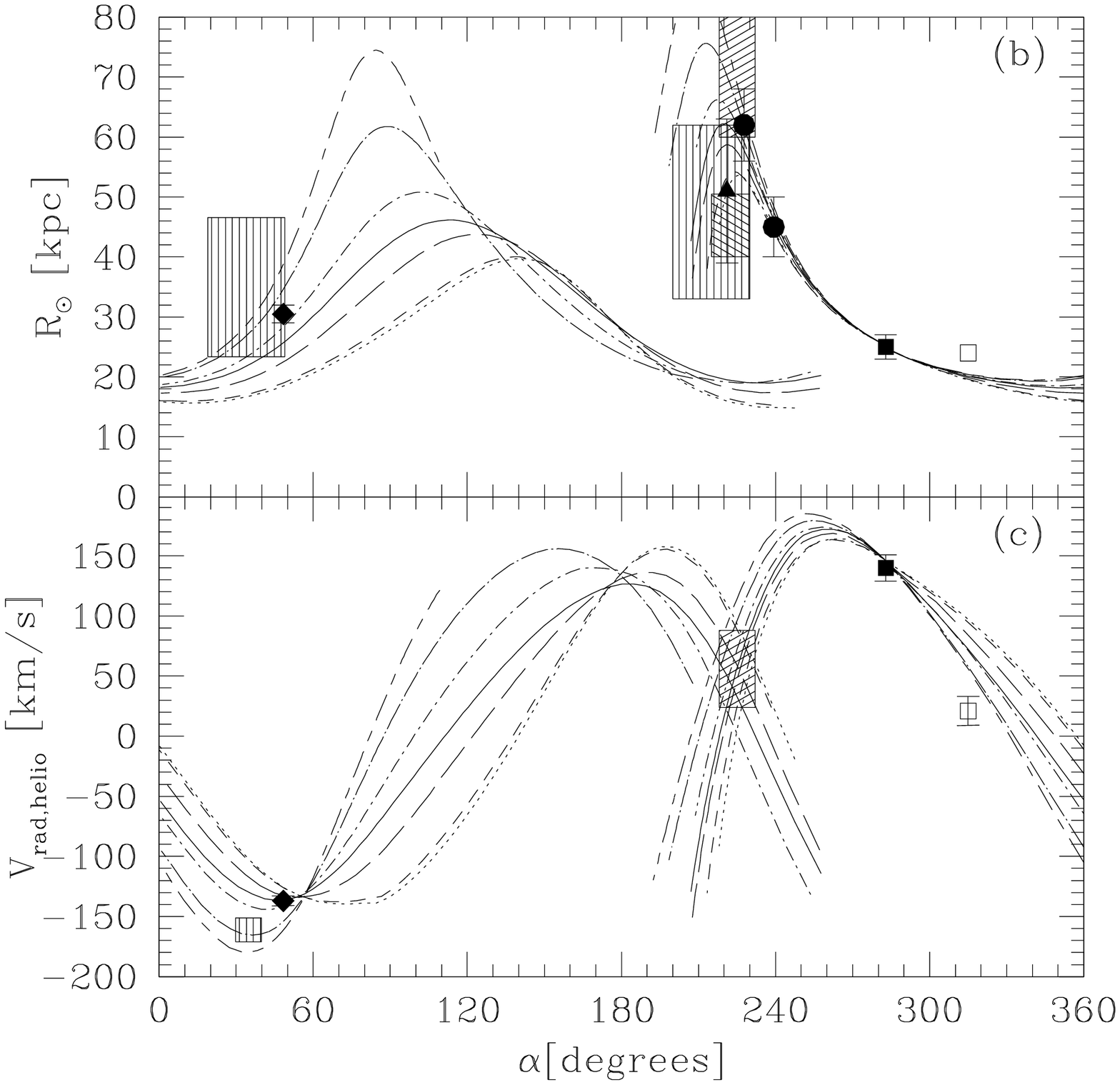}
\caption{(a) Projected position on the sky,
(b) heliocentric distances and (c) radial velocities of
the observational data and the analytical orbits of Sgr for several
flatness of the MW halo density, $q_d$.}
\end{figure}

We have compared the different Sgr orbits obtained for different
halo density flatness with the observational data of the Sgr stream.
Fig. 1a shows the projected position on the sky in
equatorial coordinates of the observational data and the analytical
orbits of Sgr. As can be seen, all the orbits have roughly the
same projected coordinates. Therefore, it is almost impossible
to infer the MW flatness using only the projection on the sky.
However, the heliocentric distances and the radial velocities provide
more information about the MW potential shape (see Fig. 1b and
1c). The heliocentric distances at the apocenter depends
highly on the density flatness. The same behavior can be observed in
the maximum radial velocity of the orbit.
From this figure we can conclude that the halo density flatness is
between 0.4 and 0.7.

\section{The potential of the Milky Way}

The physical quantity that governs the satellite orbit is
the MW potential, not the density distribution. Different density
distributions can produce almost the same MW potential and,
therefore, similar satellite orbits. However, different
potentials will always produce different
orbits. To analyse properly the results on the
shape of the MW halo from other authors, we have to
compare the MW potential flatness, not the density flatness.

Sgr spends most of its life in the outskirts of the halo (40-50 kpc).
At that distances, the density flatness inferred from the
is $q_d=0.4-0.7$, and the potential flatness is
$q_p=0.8-0.9$ (see Mart\'{\i}nez-Delgado et al 2004 for details).

These results are consistent with the MW potential flatness obtained by
other authors (Olling \& Merrifield 2000; Chen 2001; Ibata et al.
2001b), and they are also consistent with cosmological cold dark matter
models that predict oblated dark matter halos (e.g. Dubinski 1994).

\section{Numerical simulations}

We have also run numerical simulations of the Sgr satellite and the MW.
Both galaxies have been modelled using N-body systems. The simulations
have been restricted to the last few Gyrs.
During this period the MW potential has remained almost constant.
The earlier evolution of Sgr have been estimated using a quasi-equilibrium
King-Michie model (see G\'omez-Flechoso \& Dom\'{\i}nguez-Tenreiro 2001 for
details).

The flatness of the MW density distribution in the numerical
model is $q_d=0.5$, that corresponds to the better Sgr orbit obtained
in the previous section.
The characteristics of the satellite and its orbit have been chosen
forcing the model to fit the observational data of Sgr at the final
snapshot.
A detailed description of the models is
given in Mart\'{\i}nez-Delgado et al. 2004.

\subsection{Numerical results and discussion}

\begin{figure}
\plotone{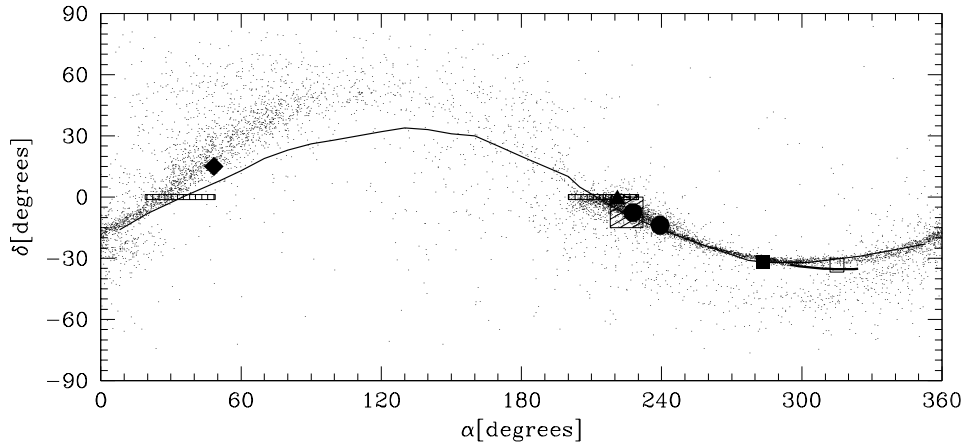}
\caption{Projected position on the sky of the numerical model of
Sgr (black dots) and the observational data (the same as Fig. 1a).}
\end{figure}

The agreement between the numerical simulations and the observational
data is very remarkable (see Fig. 2 for the
agreement in positions). Only a little disagreement in positions between these
results and the Majewski et al. stream exists in the less populated
region of the stream. This region also corresponds to the oldest part
of the stream (it has been unbound more than 4 Gyrs ago).
One reason of the disagreement could be that the MW potential
has evolved during the last 4 Gyrs and, therefore, the cosmological evolution
of the MW should be considered in order to obtain realistic old
tidal streams.
Another reason could be the uncertainty on the observational proper
motions of Sgr, since a small variation of the proper motions of the
satellite center produces large differences in the positions of the stream
at large distances of the Sgr main body.
Obviously, another reason could be the necessity of improving the MW model.

\section{Conclusions}


1. The satellite orbits can be used to constrain
the main galaxy potential.

2. Both distances and heliocentric velocities of the tidal stream are
needed to estimate the potential flatness. Using this method, we can only
infer the potential well of the main galaxy and not its mass distribution.

3. The results obtained from the Sgr stream are consistent with a MW
halo potential flatness $q_p=0.8-0.9$ at the averaged radius of the Sgr
orbit.



\begin{references}
\reference Chen, B., 2001 \apj, 553, 184
\reference Dinescu, D.I., et al. 2002, \apj, 575, L67
\reference Dubinski, J. 1994, \apj, 431, 617
\reference G\'omez-Flechoso, M.A., Fux, R., \& Martinet, L. 1999, \aap, 347, 77
\reference G\'omez-Flechoso, M.A. \& Dom\'{\i}nguez-Tenreiro, R. 2001, \apj, 550, 703
\reference Ibata, R., Gilmore, G., \& Irwin, M.J. 1994, Nature, 370, 194
\reference Ibata, R., Irwin, M., Lewis, G.F., \& Stolte, A. 2001a, \apj, 547, L133
\reference Ibata, R., et al. 2001b, \apj, 551, 294
\reference Ivezi\'c, Z., et al. 2000, \aj, 120, 963
\reference Majewski, S.R., et al. 1999, \aj, 118, 1709
\reference Majewski, S.R., Skrutskie, M.F., Weinberg, M.D., \& Ostheimer, J.C. 2003, \apj, 599, 1082
\reference Mart\'{\i}nez-Delgado, D., Aparicio, A., G\'omez-Flechoso, M.A., \& Carrera, R.  2001, \apj, 549, L199
\reference Mart\'{\i}nez-Delgado, D., G\'omez-Flechoso, M.A., Aparicio, A., \& Carrrera, R. 2004, \apj, 601, 242
\reference Mateo, M., Olszewski, E.W., \& Morrison, H.L. 1998, \apj, 508, L55
\reference Olling, R.P., \& Merrifield, M.R. 2002, \mnras, 311, 361
\reference Yanny, B., et al. 2000, \apj, 540, 825
\end{references}
\end{document}